\documentclass{vgtc}                          % final (conference style)
\ifpdf%                                % if we use pdflatex
  \pdfoutput=1\relax                   % create PDFs from pdfLaTeX
  \usepackage{graphicx}                % allow us to embed graphics files
  \DeclareGraphicsExtensions{.pdf,.png,.jpg,.jpeg} % for pdflatex we expect .pdf, .png, or .jpg files
\else%                                 % else we use pure latex
  \usepackage{graphicx}                % allow us to embed graphics files
  \DeclareGraphicsExtensions{.eps}     % for pure latex we expect eps files
\fi%

\usepackage[T1]{fontenc}% to use more chars & support copy & paste of special chars from PDF

\graphicspath{{figures/}{pictures/}{images/}{./}} % where to search for the images

\usepackage{microtype}                 % use micro-typography (slightly more compact, better to read)
\PassOptionsToPackage{warn}{textcomp}  % to address font issues with \textrightarrow
\usepackage{textcomp}                  % use better special symbols
\usepackage{mathptmx}                  % use matching math font
\usepackage{times}                     % we use Times as the main font
\usepackage{cite}                      % needed to automatically sort the references
\usepackage{tabu}                      % only used for the table example
\usepackage{booktabs}                  % only used for the table example
\usepackage{color}
\usepackage{tikz}
\usepackage{xcolor, soulutf8}
\usepackage[colorinlistoftodos,prependcaption,textsize=tiny,textwidth=1.3cm,bordercolor=orange,backgroundcolor=white,disable]{todonotes}% add 'disable' option to hide them all
\usetikzlibrary{shapes}
\usepackage{enumitem}% to remove spacing between lines
\usepackage{hyperref}
\usepackage{cleveref}
\usepackage{booktabs}
\usepackage{verbatim}
\usepackage{balance}

\DeclareRobustCommand{\colorTriangle}[1]{\tikz \draw [fill={#1},draw=none] (0,0) -- (60:.8em) -- (.8em,0) -- cycle;}

\DeclareRobustCommand{\threeCircles}[1]{
\tikz \draw [fill={#1},draw=none] circle (.18em);
\tikz \draw [fill={#1},draw=none] circle (.18em);
\tikz \draw [fill={#1},draw=none] circle (.18em);
}

\definecolor{colorEvent}{HTML}{5b85e2}
\definecolor{colorNormal}{HTML}{82D6C3}
\definecolor{colorGray}{HTML}{797979}
\newcommand{\gray}[1]{\textcolor{colorGray}{#1}}
\newcommand{\filteredData}{\gray{\small \faStar}}
\newcommand{\unfilteredData}{\gray{\small \faStarO}}

\newcommand{\REMOVE}[1]{\ignorespaces}
\newcommand{\ADD}[1]{#1}

\onlineid{0}

\vgtccategory{Research}

\vgtcinsertpkg

\preprinttext{Original submission to IEEE VIS short papers 2019}

\title{Evaluating Alignment Approaches in Superimposed Time-Series and Temporal Event-Sequence Visualizations}

\newcommand{\orcidicon}[2]{%
  \href{#2}{#1 \includegraphics[height=\fontcharht\font`\B]{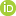}}%
}

\author{
    \orcidicon{Yixuan Zhang}{https://orcid.org/0000-0002-7412-4669}%
    \thanks{%
        Corresponding author.
        E-mails: [~%
        \href{mailto:zhang.yixua@husky.neu.edu}{zhang.yixua} |
        \href{mailto:dibartolomeo.s@husky.neu.edu}{dibartolomeo.s} |
        \href{mailto:sheng.f@husky.neu.edu}{sheng.f}~]@husky.neu.edu, 
        [~%
        \href{mailto:h.jimison@northeastern.edu}{h.jimison} | 
        \href{mailto:c.dunne@northeastern.edu}{c.dunne}~]@northeastern.edu
        \vspace{0.55in}
    }\\%
    \and \orcidicon{Sara Di Bartolomeo}{https://orcid.org/0000-0001-9517-3526}\\%
    \and \orcidicon{Fangfang Sheng}{https://orcid.org/0000-0002-2101-3754}\\%
    \and \orcidicon{Holly Jimison}{http://orcid.org/0000-0003-2374-382X}\\%
    \and \orcidicon{Cody Dunne}{https://orcid.org/0000-0002-1609-9776}\\%
}
\affiliation{\scriptsize Northeastern University}

\teaser{
    \centering
    \includegraphics[width=\linewidth]{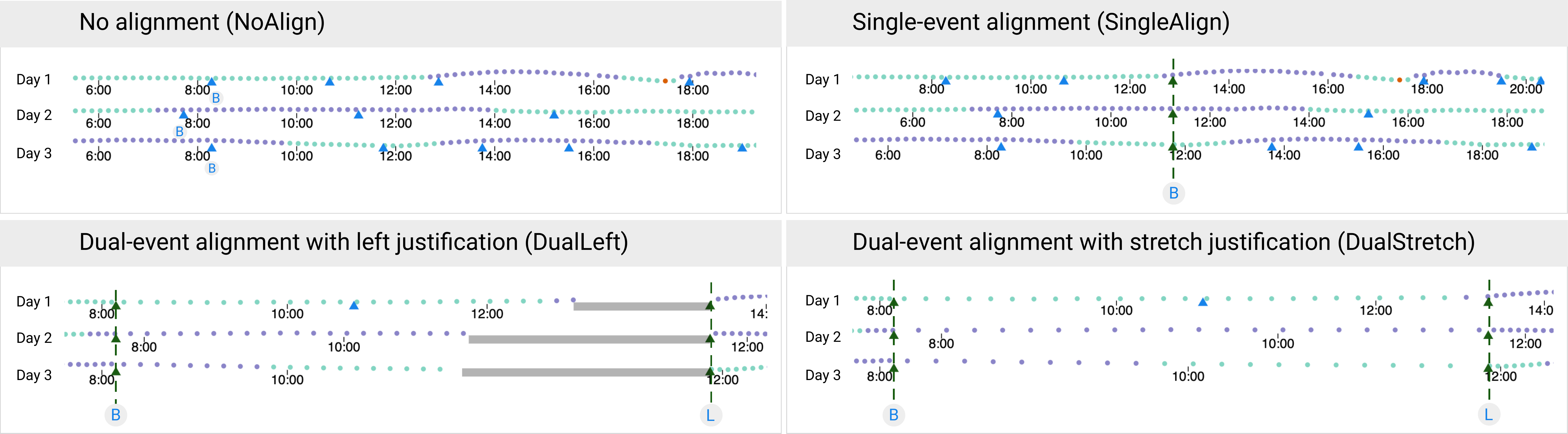} 
    \caption{
    User task performance was evaluated on four temporal event sequence alignment visualization approaches: no sentinel event alignment (NoAlign), single-event alignment (SingleAlign), dual-event alignment with left justification (DualLeft), and dual-event alignment with stretch justification (DualStretch).
    Stimuli were created using IDMVis \cite{zhang2019idmvis} and Type 1 diabetes treatment data.
    Measures were task completion time, correctness, and error rate.
    }
    \label{fig:teaser}
}

\shortauthortitle{Zhang \MakeLowercase{\textit{et al.}}: Evaluating Alignment Approaches in Superimposed Time-series and Temporal Event-sequence Visualizations}
\abstract{
Composite temporal event sequence visualizations have included sentinel event alignment techniques to cope with data volume and variety.
Prior work has demonstrated the utility of using single-event alignment for understanding the precursor, co-occurring, and aftereffect events surrounding a sentinel event. However, the usefulness of single-event alignment has not been sufficiently evaluated in composite visualizations.
Furthermore, recently proposed dual-event alignment techniques have not been empirically evaluated. In this work, we designed tasks around temporal event sequence and timing analysis and conducted a controlled experiment on Amazon Mechanical Turk to examine four sentinel event alignment approaches: no sentinel event alignment (NoAlign), single-event alignment (SingleAlign), dual-event alignment with left justification (DualLeft), and dual-event alignment with stretch justification (DualStretch). Differences between approaches were most pronounced with more rows of data. For understanding intermediate events between two sentinel events, dual-event alignment was the clear winner for correctness---71\% vs. 18\% for NoAlign and SingleAlign. For understanding the duration between two sentinel events, NoAlign was the clear winner: correctness---88\% vs. 36\% for DualStretch---completion time---55 seconds vs. 101 seconds for DualLeft---and error---1.5\% vs. 8.4\% for DualStretch. For understanding precursor and aftereffect events, there was no significant difference among approaches.
A free copy of this paper, the evaluation stimuli and data, and source code are available at \href{https://osf.io/78fs5/}{\texttt{osf.io/78fs5}}
}

\CCScatlist{
  \CCScatTwelve{Human-centered computing}{Visu\-al\-iza\-tion}{Empirical studies in visualization}{} 
}

\nocopyrightspace

\begin{document}


\firstsection{Introduction}

\maketitle

Time-series data visualization has been widely used in a variety of fields, including healthcare, business, engineering, and social media because of its ability to show how variables change over time.
In recent years, visualizing discrete events on a time axis (temporal event sequence visualization) is also gaining  popularity as this approach can reveal event patterns over time. 
Recent studies have started to composite these two types of visualization by overlaying the event sequence on top of time-series data visualization. 
As the superimposed view highlights the relationship between the events and the time-series data, the superimposition demonstrates the great potential to facilitate the process of making sense of the data~\cite{javed2012exploring}. 

Among various types of strategies to tackle the growing volume and complexity of data, alignment is a commonly adopted approach. 
With alignment, a visualization tool can put multiple instances of the same type of events at the same vertical or horizontal position, so that users can easily see the interaction between the aligned events and other events. 
In superimposed time-series and event-sequence visualizations, alignment can help discover the precursor (what triggers the event) and the aftereffect (what the event triggers) of the event,  
\ADD{in the same visual space. Appropriately using superimposition would help ease comparison across the time-series data and related attributes displayed in the event-sequence visualization. However, applying alignment techniques in superimposed views might also increase more visual clutter and cognitive load.}
%\ADD{However, superimposition of views are more likely to result in cognitive load and visual clutter~\cite{javed2012exploring, beecham2016faceted}.}
The usefulness of aligning by one sentinel event has been explored by Wang et al.~\cite{wang2008aligning}.
However, previous evaluations focus on temporal event-sequence visualizations, while alignment techniques in superimposed visualizations are not sufficiently studied.

Moreover, the visualization community is no longer satisfied with single-event alignment (e.g.,~\cite{wang2008aligning}), and dual-event alignment~\cite{zhang2019idmvis} has been proposed to explore the more complex temporal data interaction. 
IDMVis~\cite{zhang2019idmvis} exploits three different types of dual-event alignment to visualize the interaction of more than one event with the underlying value; the tool has received positive feedback from domain experts.
However, the evaluation with domain experts cannot be easily generalized. A formal study that compares the effectiveness of single-event and dual-event alignment is required before widely accepting the single-event alignment and dual-event alignment in superimposed time-series and event-sequence visualizations.
Therefore, we designed a series of tasks that are tailored for the superimposed time-series and temporal event-sequence visualization and a controlled experiment to evaluate different temporal event alignment techniques, including no  alignment, single-event alignment, and dual-event alignment.
%two variants of dual-event alignment (with left justification and stretch justification).

\section{Related Work} 

\subsection{Temporal Event Sequence Visualization}
Visualization researchers have proposed a variety of techniques to deal with temporal event sequences, such as temporal folding and aligning. 
Temporal event alignment is a strategy to reduce pattern complexity by ``aligning all records by the occurrence of a selected alignment''~\cite{du2017coping}.
The purpose of applying aligning techniques is to help identify precursor, co-occurring, and aftereffect events~\cite{wang2008aligning}. 
Alignment techniques have been widely used in different domains, such as healthcare~\cite{gschwandtner2011carecruiser, wang2008aligning}, application log analysis~\cite{chen2018sequence}, and sports~\cite{monroe2013temporal}.
For example, %in the healthcare field,
LifeLines2~\cite{wang2008aligning} allowed users to align, rank, and filter medical events, such as first heart attack. 
In the application log analysis case study~\cite{chen2018sequence}, alignment enabled analysts to identify the antecedents of the aligned events when the error messages appeared.
Alignment can also be used to support cause-and-effect relation analysis. 
A similar case of analyzing Twitter user activity logs allowed users to pick a customized alignment point of interest---by default the beginning of events---to identify the sequence before and after the alignment point~\cite{wongsuphasawat2014using}.  
In these prior works, event alignment can be interacted with using a single view.  

Other researchers explored the design space of applying alignment in composite views. Using alignment techniques, CareCruiser~\cite{gschwandtner2011carecruiser} supported comparison between the effects of the selected treatment plan among different patients as well as evaluating alternative treatment plans for a single patient. 
In our previous work by Zhang et al., we introduced the dual-event alignment in IDMVis~\cite{zhang2019idmvis} to help clinicians make treatment decisions for patients with Type 1 diabetes. Though our evaluation with domain experts was positive and gave us ample feedback, there has been a lack of thorough evaluation of the usefulness of using dual-event alignment versus single-event alignment. In this work, we aim to understand the effectiveness of different event alignment approaches in superimposed time-series and temporal event-sequence visualizations.

\subsection{Tasks in Understanding Temporal Event Sequences}
\label{subsec:tasks}
Temporal event-sequence visualizations can be used to support a wide range of sequence and timing analysis tasks. 
Along with popularizing the idea of sentinel event alignment for event-sequence analysis, the authors of Lifelines2~\cite{wang2008aligning} described a set of six low-level domain-specific tasks.
Here we re-frame these as three general tasks and present the results of their controlled experiment. 
Task~1: Given an event of interest and a subsequent time window, count aftereffect events of given event type within that window.
In this case, participants viewing 5 records performed similarly in the non-aligned and aligned conditions.
However, when viewing 20 records participants were on average 65\% faster, more accurate, and had less error.
Task~2: Given an event of interest and a subsequent time window, explore the most frequent aftereffect or co-occurrence events within that window.
In one case, participants were on average 62\% faster but there were no differences in accuracy or error.
In others, the results were mixed.
Task~3: Given \textit{two} events of interest, count intermediate events that match a given event type.
In this case participants were possibly more accurate but there were no differences in time or error.

The authors of DecisionFlow~\cite{gotz2014decisionflow} described 12 high-level tasks focusing on three dimensions: interpretation of the temporal flow, understanding the properties of intermediate events, and reasoning about the changes of event sequences.
Among these 12 tasks, those most related to sequence and timing analysis were (1) understanding temporal flow relationships (e.g., A before B) and (2) sentinel event (milestone) time comparisons.
Plaisant and Shneiderman~\cite{plaisant2016diversity} further developed eight high-level tasks for event analytics across three high-level categories: heightening awareness, identifying data for further study, and understanding the impact of event patterns.
This last category includes a particularly relevant task for sequence and timing analysis that harkens back to Lifelines2~\cite{wang2008aligning} tasks: given an event of interest and optionally a surrounding time window, characterize the precursor (antecedent) or aftereffect (sequelae) events.
However, most of the tasks prior researchers have considered have been too high-level, domain-specific, or unrelated to sequence and timing analysis to be used as guidance for developing tasks for our evaluation.
Therefore, we first must build domain-independent task abstractions for temporal event sequence and timing visualization. 

\section{Hypotheses}
 
The purpose of this study is to understand the effectiveness of event-alignment in superimposed time-series and temporal event-sequence visualizations. We form the hypotheses as follows:

\textbf{H1:} 
\textit{The benefit of using single event alignment will be greater than no alignment for identifying precursor and aftereffect events and groups of events.}
Rationale: This hypothesis is grounded in prior works (e.g., Lifelines2~\cite{wang2008aligning}) which show that aligning by a sentinel event improved completion time and accuracy.
We need to check if the previous results still hold in superimposed visualizations. 

\textbf{H2:} 
\textit{There is no benefit of using single- vs. dual-event alignment for identifying precursor and aftereffect events.} %and groups of events. 
Rationale: Since dual-event alignment builds upon single-event alignment and is similar for a single sentinel event, we assume they perform similarly. 

\textbf{H3:} 
\textit{Dual-event alignment will be more useful than single-event alignment or no alignment for understanding intermediate events.}
Rationale: This hypothesis is based on the IDMVis qualitative evaluation with domain experts which indicates that using dual-event alignment will increase performance~\cite{zhang2019idmvis}.

\section{Comparative Evaluation}

We conducted a controlled experiment on Amazon Mechanical Turk (MTurk) to assess the effect of sentinel event alignment approaches on participant task performance.

\subsection{Stimuli and Tasks}

\begin{figure}[tb]
 \centering % avoid the use of \begin{center}...\end{center} and use \centering instead (more compact)
 \includegraphics[width=\columnwidth]{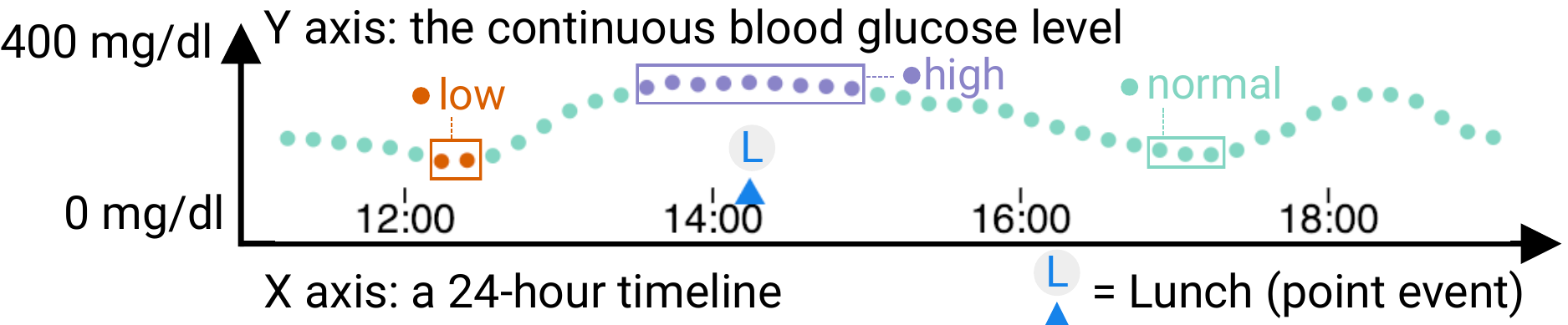}
 \vspace{-20pt}
 \caption{Legend of the main visual encodings used in the study.}
 \vspace{-20pt}
 \label{fig:legend} 
\end{figure}

\begin{figure*}[tb]
 \centering % avoid the use of \begin{center}...\end{center} and use \centering instead (more compact)
 \includegraphics[width=\textwidth]{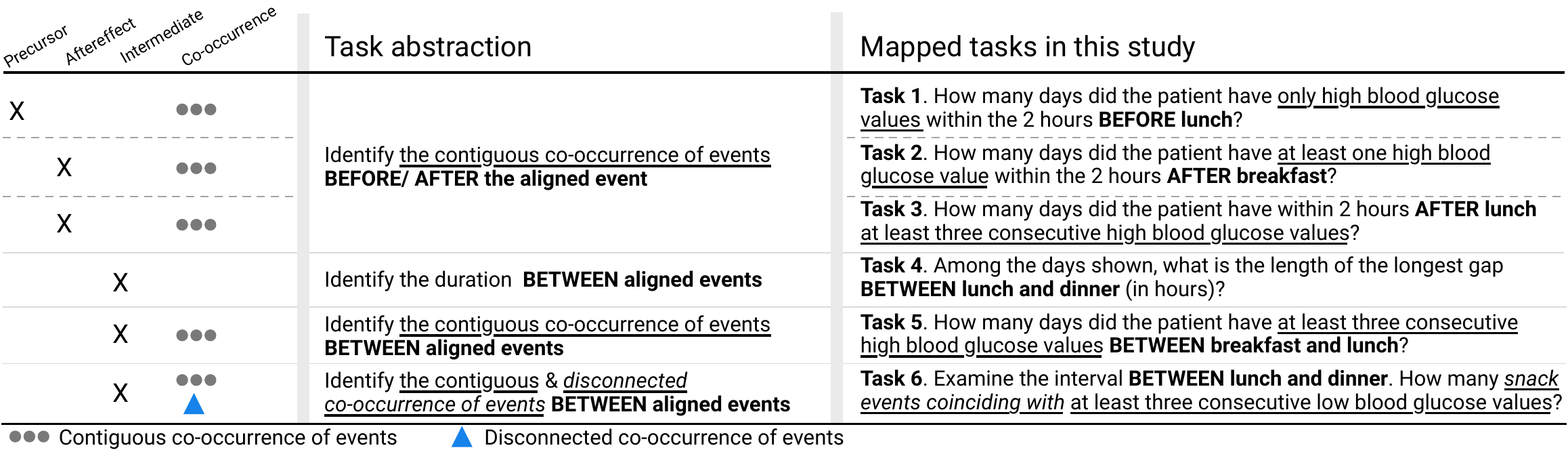}
 \vspace{-18pt}
 \caption{Tasks for evaluating precursor, aftereffect, intermediate, and co-occurrence of events in superimposed %time-series and temporal event-sequence 
 visualizations.}
 \vspace{-20pt}
 \label{fig:tasks}
\end{figure*}

We selected stimuli from our open source tool IDMVis~\cite{zhang2019idmvis} using its de-identified Type 1 diabetes treatment data.
Example stimuli are shown in \autoref{fig:teaser} for the four alignment approaches.
The main visual encodings are described in \autoref{fig:legend}, including 
point events (e.g., lunches) shown as labeled blue triangles \colorTriangle{colorEvent};
time series data for a continuous glucose monitor shown as color-coded dots \threeCircles{colorNormal}; and the time and value axes.
We selected tasks (detailed in~\autoref{fig:tasks}) guided by our hypotheses,  a synthesis of prior work in temporal event sequences discussed in~\autoref{subsec:tasks}, and taskonomies for Type 1 diabetes management \cite{zhang2019idmvis, bergenstal2013recommendations}. 
\ADD{We designed a set of low-level tasks for evaluation considering the insufficiency of high-level tasks for sensemaking in the clinical context.
In chronic illness management, sensemaking requires iterative data exploration and feature selection before formalizing hypothesis formulation and evaluation~\cite{mamykina2017personal}.}

To examine co-occurrence of events, we categorize co-occurrence of events into \colorTriangle{colorEvent} disconnected and \threeCircles{colorGray} contiguous co-occurrence of events. 
Contiguous co-occurrence of events are events that are adjacent to each other. Such events can include ``interval events'' (e.g., exercise events with duration logged) but also time series data (e.g., continuous blood glucose level) as the data points are contiguous.
Disconnected co-occurrence of events are events that do not have constraints on adjacency.
They are similar to the concept of ``point events'' (e.g., insulin injection).  We designed questions to understand both \colorTriangle{colorEvent} disconnected (Task 6) and \threeCircles{colorGray} contiguous co-occurrence of events (Task 1,2,3,5,6).

We are also interested in assessing participant task performance at various visualization scales.
Our visualization stimuli consist of one row for every day of data.
We use a small scale of three days (as shown in
\autoref{fig:teaser}) as well as a larger scale of 14 days.
We asked each participant to answer 12 questions: 6 composite tasks$\times$2 scales. Questions 1--6 used 3-day data visualizations and Questions 7--12 used 14-day data visualizations. In total, we selected a disjoint subset of the temporal event sequence for each question and created 48 visualization stimuli: 6 composite tasks$\times$2 scales$\times$4 conditions.
To avoid ordering effects we randomized the order in which we presented the tasks to each participant.

% The six tasks and the design rationale are discussed below:
% Task 1: How many days did the patient have only high blood glucose values within the 2 hours before lunch? 

\subsection{Procedure}
In this study we used a between-subjects approach to separate participants randomly but equally between four experimental conditions (shown in~\autoref{fig:teaser}): no sentinel event alignment (NoAlign), single-event alignment (SingleAlign), dual-event alignment with left justification (DualLeft), and dual-event alignment with stretch justification (DualStretch). 
We did not test dual-event alignment with right justification as we believe that participant performance would mirror dual-event alignment with left justification.

Upon approval from the Institutional Review Board of our institution, we recruited participants from MTurk. 
Informed consent was gained before they accepted the HIT. 
Participants were asked to read through a brief introduction to the domain background (Type 1 diabetes) and a tutorial with example questions, answer 12 questions about alignment visualizations, and answer a questionnaire about their experience using alignment approaches and their demographics. 
Participants completed the study in 12 minutes 59 seconds on average.
They were compensated with a base pay of \$4 USD---average hourly rate \$18.48---with a bonus of \$0.60 for every correct answer after the sixth one.
Participants were given a 10\% and 20\% bonus if they completed the study in less than 12 minutes and 10 minutes, respectively. 
\REMOVE{The purpose of providing a bonus is to encourage fast and correct performance.}
\ADD{We provide bonus rewards to encourage both fast and correct performance, as discussed in \cref{sec:discussion}.}
To ensure honest participation and sufficient understanding of the study and tasks, based on the pilot results we set the rejection criteria as participants (1) having less than or equal to one correct answer AND spending less than eight minutes in total, OR (2) less than one minute on the tutorial, OR (3) less than four minutes on questions.

\subsection{Participants} 
\label{subsec:participants}
Based on mean and variance estimates from a pilot study, we conducted a power analysis---Type I error rate $\alpha = 0.05$, power $1-\beta = 0.8$---from which we determined we needed 108 participants.
In total we recruited 123 U.S. participants and accepted 108. \REMOVE{ (42 female, 76 male, 1 non-binary, and 4 prefer not to say) on MTurk.}
After applying a k-means clustering of the time spent on the questions and the numbers of correct answers to all accepted participants, we discovered that there was a clear clustering effect and identified 46 participants as ``speeding'' with the bound as four answers out of 12 questions and at least 8 minutes were spent on the questions. Therefore, we filtered out 46 participants leaving us 62 \ADD{(16 for NoAlign, 18 for SingleAlign, 14 for DualLeft, and 14 for DualRight)} 
for analysis.
%Below we report results both for the filtered \filteredData~(n=62) and unfiltered \unfilteredData \ data~(n=123).
Among the 62 participants, the median age was 34 ($IQR = 15.5$). 
%In terms of educational attainment, 
%21 participants had a Bachelor's degree, 14 had a high school Degree, 13 had college experience but no degree, 10 had an Associate's degree, and 3 had a Master's degree; one participant did not disclose the information. 

\subsection{Measures and Analysis}
Due to the composite nature of our questions, we analyzed participant performance separately for each question.
For each question, we measured the completion time, correctness, and error rate. 
The completion time is the time, in minutes, a participant spent to complete the question.
The correctness is a binary measure, measured by whether or not the participant got the right answer. If the participant answered the question correctly, correctness was recorded as 1, otherwise treated as 0. 
We report error rate ($ER$) using $ER = \mid O_p - O_t \mid / O_t $, where the $O_p$ and $O_t$ represent the option that the participant selected and the correct answer, respectively.   
To select the appropriate statistical analysis plan, we first tested for normality with two methods qualitatively and quantitatively by using Q-Q plots and the Shapiro-Wilk test respectively. 
Our results show that our data does not have a normal distribution. Therefore, we used chi-square test for independence to analyze correctness and the Kruskal-Wallis H test as the nonparametric alternative to the one-way ANOVA to analyze time and error rate. We used the Bonferroni Adjustment for post hoc tests.

\section{Results \ADD{and Discussion}}
\label{sec:results}
\begin{figure}[tb]
 \centering % avoid the use of \begin{center}...\end{center} and use \centering instead (more compact)
 \includegraphics[width=\columnwidth]{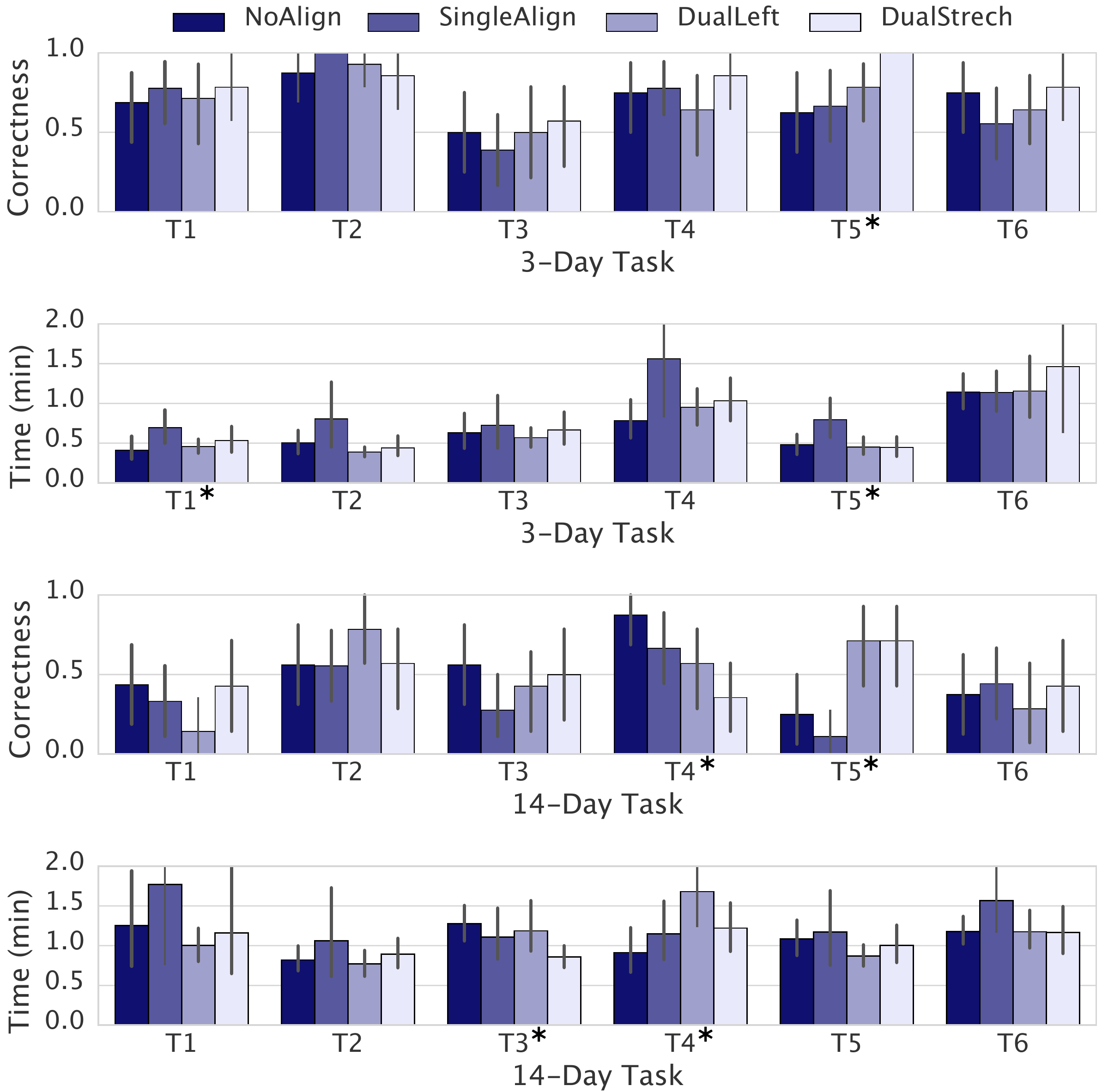}
 \vspace{-18pt}
 \caption{Correctness and time by scale, task, and approach with bootstrapped 95\% confidence intervals\ADD{, with asterisk (*) representing the results having significant~($p<.05$) differences. \REMOVE{(``speeders'' filtered out \filteredData).} 
 }}
 \label{fig:combined_chart}
 \vspace{-15pt}
\end{figure}

We present the results using the filtered data \ADD{($n=62$)} \REMOVE{(\filteredData,~$n=62$) }
which removes likely ``speeding'' actors as described in~\autoref{subsec:participants}.
\autoref{fig:combined_chart} shows correctness and completion time for each combination of scale, task, and approach.
For space considerations \REMOVE{details} \ADD{ results with the unfiltered data and charts} of error are relegated to supplemental material.
\REMOVE{Inline we mark results that were also significant in the \textit{\textbf{unfiltered data}} ($n=123$) with \unfilteredData.}

\subsection{Precursor and Aftereffect Events}
Tasks~1--3: 
We did not observe any significant differences \ADD{($p<0.05$)} between groups in terms of correctness and error rate. 
Participants were significantly slower using SingleAlign (45s) than NoAlign (31s) for the 3-day scale~($p=.02$) \REMOVE{(\filteredData~$p=.02$)}
---but not for the 14-day scale.
Contrary to our hypothesis, our results indicate that NoAlign was equal or better than SingleAlign for identifying precursor and aftereffect events.
These results seem to contradict the Lifelines2~\cite{wang2008aligning} evaluation which showed an improvement in completion time.

However, \ADD{there are several differences we must point out.
%We evaluate the visual representations provided by IDMVis~\cite{zhang2019idmvis}, which was designed for analyzing Type 1 diabetes treatment data.
}
The Lifelines2~\cite{wang2008aligning} study explored point events, whereas we examine more complex superimposed visualizations of multiple data sources. 
\ADD{%the nature of the complexity of superimposed visualization itself might also have an impact on the completion time and accuracy. 
Superimposed visualizations allow direct comparison in the same visual space, but may also cause high visual clutter and thus sacrifice readability~\cite{javed2012exploring}.
We also visualize event types differently than LifeLines2.
%In addition to displaying meal events using triangles, we also treat blood glucose levels as point events using dots that encode quantitative values using color and position.
We display point events (e.g., breakfast) similarly using triangles. However, blood glucose levels are also treated as point events shown by dots that encode quantitative values using color and position.
%The visual representation we use is just one of the many possible ways to visualize time-series and temporal event-sequence visualizations. 
The complex Type 1 diabetes data and associated domain tasks~\cite{zhang2019idmvis} may have also contributed to task difficulty.
Tasks that involve understanding contiguous co-occurrence events were complex, but have parallels with the tasks used to evaluate LifeLines2.
%Tasks that involve understanding contiguous co-occurrence events (e.g., ``three consecutive high blood glucose levels'') were complex, but have parallels with the tasks used to evaluate LifeLines2.
}

\REMOVE{On the other hand, our results support our hypothesis that single- and dual-event alignment approaches are equivalent for identifying precursor and aftereffect events.}

\subsection{Intermediate Events}
Task~4:
For the 3-day scale we did not observe any significant differences in terms of completion time, correctness, and error rate.
For the 14-day scale however, participants were more correct using NoAlign (88\%) than DualStretch (36\%)~\REMOVE{(\filteredData~$\chi^2 = 6.54$, $p = .01$; \unfilteredData~$\chi^2 = 5.93$, $p = .02$)}\ADD{($\chi^2 = 6.54$, $p = .01$)}, were faster using NoAlign (55s) than DualLeft (101s)~\REMOVE{(\filteredData~$p = .02$)}\ADD{($p = .02$)}, and had less error using NoAlign (2\%) than DualStretch (8\%)~\REMOVE{(\filteredData~$p = .00$; \unfilteredData~$p = .03$)}\ADD{($p = .00$)}.
This suggests that dual-event alignment is not an appropriate solution for identifying interval duration.

Task~5:
For the 3-day scale, participants were faster using NoAlign (47s) than SingleAlign (94s)~\REMOVE{(\filteredData~$p = .05$)}\ADD{($p = .05$)}, more correct using DualStretch (100\%) than NoAlign (62\%)~\REMOVE{(\filteredData~$\chi^2 = 4.43$, $p = .04$)}\ADD{($\chi^2 = 4.43$, $p = .04$)}, and had less error using DualStretch (0\%) than NoAlign (14\%)~\REMOVE{(\filteredData~$p = .01$)}\ADD{($p = .01$)}.
For the 14-day scale, we did not observe any significant difference in terms of completion time.
However, participants were more correct using DualStretch (71\%) and 
DualLeft (71\%) than NoAlign (25\%)~\REMOVE{(\filteredData~$\chi^2 = 4.74$, $p = .03$)}\ADD{($\chi^2 = 4.74$, $p = .03$)} or SingleAlign (11\%)~\REMOVE{(\filteredData~$\chi^2 = 9.79$, $ p = .00$)}\ADD{($\chi^2 = 9.79$, $ p = .00$)}.
These results support our hypothesis that dual-event alignment will be beneficial for understanding intermediate events, as long as they are not based on interval duration (Task~4).

Task~6: 
We did not observe any significant difference in terms of completion time, correctness, and error rate.
One possible explanation from the participants was that the wording of Task 6 was unclear.
For example, %at the end of the study, 
one participant commented that \textit{``[the] wording was confusing, because I thought it meant that both of those things happened between the two meals, not simultaneously, so I got that one wrong unfortunately.''}
However, we had not observed any difficulties in interpretation of the tasks in our pilot studies. %working with visualization and clinical colleagues as well as on MTurk .
One important take-away from this study is that the design of user study tasks requires substantial design and usability iteration.

% ---- for supplementary material ---- 
%We observed a few differences between using the filtered data and non-filtered data. We report the differences by tasks using non-filtered data.
%Task 4:  there was no significant difference between using NoAlign and DualLeft in terms of completion time. 
%Task 5: There was a significant difference between using DualStretch and DualLeft in terms of correctness.
 
\section{\REMOVE{Discussion }\ADD{Reflections and Design Implications}}
\label{sec:discussion}
In this controlled experiment on MTurk, we measured task completion time, correctness, and error rate to evaluate the effect of different sentinel event alignment techniques in superimposed time-series and temporal event-sequence visualizations. 
\ADD{
We highlight several design implications for using alignment techniques, as well as design considerations when designing relevant tasks for evaluation.}

\ADD{First, consider using dual-event alignment for exploring intermediate events, regardless of the visualization scale (e.g., number of days). 
In most cases, DualStretch performed slightly better than DualLeft. 
However, dual-event alignment might not be most suitable for identifying the duration between two events.}

\ADD{Second, %a few design considerations appear when designing for event sequence visualizations using alignment techniques, based on the discussion of the possible confounding factors related to the complexity of tasks in ~\autoref{sec:results}. 
when applying alignment techniques, the nature of the composite visualizations should be taken into consideration.
The complexity of a superimposed time-series and event-sequence visualizations is likely to add cognitive burdens to users.  
We need to further evaluate the design space of superimposed time-series and temporal event-sequence visualizations such as showing categorical tags or types of events through alignment alone, using categorical shape, color, and position.
Also, we believe it is necessary to perform a more systemic evaluation on temporal event sequence in composite visualization that also includes juxtaposed, integrated, overloaded, and nested views---in addition to superimposed visualizations~\cite{javed2012exploring}.} 

\ADD{Third, we suggest a call for action for the design of low-level tasks.
We noted that the complexity of the low-level tasks related to superimposed %time-series and temporal event-sequence 
visualizations might affect users' performance. 
%Though we believe these tasks may be transferable to domains outside of the diabetes treatment example, there are opportunities to expand on these tasks to encompass the wide variety of other goals analysts may have (e.g., \cite{plaisant2016diversity}). 
When designing tasks for non-experts on crowdsourcing platforms, the appropriate balance between workload, ease of tasks, and domain knowledge remains unclear and thus requires further study.
%When designing tasks for non-experts, especially on crowdsourcing platforms (e.g., MTurk participants), the appropriate balance between workload, ease of the tasks, and domain knowledge remains unclear and thus requires further study.
}

\ADD{Fourth, participant compensation requires thoughtful consideration.
The monetary reward is the primary motivator for many workers~\cite{chen2011opportunities} who may have faster response times than in lab studies due to optimizing money/time~\cite{Kosara2010DoMechanicalTurks}.
Performance-based payments, e.g. a correctness bonus, may increase work quality but also increase the amount of time workers spend~\cite{Ho2015IncentivisingHighQualityCrowdwork}.
In order to use time as one of our measures, we use a speed bonus to help counteract any increase in time spent.
Separate bonuses could lead to a situation in which a participant favors optimizing one over the other.
We did not detect any negative impact of our bonus setup but it merits further study.}

%\section{\ADD{Limitations}}

\section{Conclusions}
We designed low-level tasks for superimposed time-series and event-sequence visualizations, and
\REMOVE{ these tasks may be used as an inspiration tool for future research. We} 
conducted a comparative evaluation on MTurk to understand the effect of different sentinel event alignment approaches.
\REMOVE{in superimposed time-series and temporal event-sequence visualizations.}
We found that dual-event alignment allows users to perform tasks of intermediate events more accurately and quickly, except for identifying the duration.
The benefits of using single-event alignment were not significant, compared with no alignment. 
There is no significant difference in using single- vs. dual-event alignment for identifying precursor and aftereffect events in terms of task completion time, correctness, and error rate.  
We encourage visualization researchers to conduct further studies to assess the effect of alignment techniques in composite visualizations. 

\acknowledgments{We thank the National Science Foundation for support under CRII award no. 1755901 and our reviewers \& colleagues for their advice.
A special thanks to Jane Kokernak for editing and proofreading.}  

% \balance

%\bibliographystyle{abbrv}
%\bibliographystyle{abbrv-doi}
%\bibliographystyle{abbrv-doi-narrow}
\bibliographystyle{abbrv-doi-hyperref}

\bibliography{template}
\end{document}